\documentstyle[12pt]{article}
\def\simgt{\stackrel{>}{{}_\sim}}
\def\be{\begin{equation}}
\def\ee{\end{equation}}
\def\bear{\be\begin{array}}
\def\eear{\end{array}\ee}
\def\bea{\begin{eqnarray}}
\def\eea{\end{eqnarray}}

\def\baselinestretch{1}
\begin{document}
\catcode`@=11
\newtoks\@stequation
\def\subequations{\refstepcounter{equation}%
\edef\@savedequation{\the\c@equation}%
  \@stequation=\expandafter{\theequation}
  \edef\@savedtheequation{\the\@stequation}
  \edef\oldtheequation{\theequation}%
  \setcounter{equation}{0}%
  \def\theequation{\oldtheequation\alph{equation}}}
\def\endsubequations{\setcounter{equation}{\@savedequation}%
  \@stequation=\expandafter{\@savedtheequation}%
  \edef\theequation{\the\@stequation}\global\@ignoretrue

\noindent}
\catcode`@=12
\begin{titlepage}
\title{{\bf The Generalized Dilaton Supersymmetry Breaking Scenario} 
\thanks{Research supported in part by: the CICYT, under
contracts AEN95-0195; the European Union,
under contract CHRX-CT92-0004.}
}
\author{ {\bf J.A. Casas\thanks{On leave of absence from Instituto de
Estructura de la Materia CSIC, Serrano 123, 28006 Madrid, Spain.}
${}^{ {\footnotesize,\S}}$}\\
\hspace{3cm}\\
${}^{\footnotesize\S}$ {\small Santa Cruz Institute for Particle Physics}\\
{\small University of California, Santa Cruz, CA 95064, USA}}
\date{}
\maketitle
\def\baselinestretch{1.15}

\begin{abstract}
\noindent
We show that the usual dilaton dominance scenario, derived
from the tree level K\"ahler potential, can never correspond to 
a global minimum of the potential at 
$V=0$. Similarly, under very general assumptions it cannot correspond
to a local minimum either, unless a really big conspiracy of different
contributions to the superpotential $W(S)$ takes place.
These results, plus the fact that the K\"ahler potential is
likely to receive sizeable string non-perturbative contributions,
strongly suggest to consider a more general scenario, leaving the 
K\"ahler potential arbitrary. In this way we obtain generalized
expressions for the soft breaking terms but a predictive scenario still
arises. Finally, we explore the phenomenological capability of some
theoretically motivated forms for non-perturbative K\"ahler potentials,
showing that it is easy to stabilize the dilaton at the 
realistic value $S\sim 2$ with just one condensate and no fine-tuning.

\end{abstract}

\thispagestyle{empty}

\vspace{3cm}
\leftline{SCIPP-96-20}
\leftline{IEM-FT-129/96}
\leftline{May 1996}
\vskip-21.cm

\rightline{SCIPP-96-20}
\rightline{IEM-FT-129/96}

\end{titlepage}
\newpage
\setcounter{page}{1}

\section{Introduction and outline}

The so-called ``Dilaton Dominance'' \cite{Kaplu93, Brignole} 
model has become a popular scenario
of supersymmetry (SUSY) 
breaking in the framework of superstrings. The basic assumptions 
of this scenario are the following. 

\begin{enumerate}

\item Supersymmetry is
broken in the dilaton field ($S$) sector. In other words, only the $F_S$
auxiliary field is to take a non-vanishing VEV.

\item The dilaton dependence
of the K\"ahler potential, $K$, is assumed to be sufficiently well
approximated by the tree-level expression, $K=-\log(S+\bar S)$.

\item The superpotential $W$ is in such
an (unknown) way that the minimum of the potential lies at an acceptable
value for the dilaton. More precisely, since $\langle{\rm Re} S\rangle=
g_{string}^{-2}$, i.e. the unified gauge coupling constant at the
string scale, $\langle S\rangle\simeq 2$ has to be assumed.

\item The condition of vanishing cosmological constant, i.e.
$V=0$ at the minimum of the potential is also normally assumed.

\end{enumerate}

The first assumption is plausible since the dilaton field is always
present in any string scenario (at least in the weak coupling approach)
and it has Planck-mass suppressed couplings with all the matter fields,
i.e. it is a ``hidden-sector'' field. The second assumption is attractive
since the tree-level form of $K(S,\bar S)$ is universal for any string
construction, thus providing a model independent framework. The third
assumption is phenomenologically mandatory, while the last assumption
is the usual one in supergravity (SUGRA) models\footnote{For a discussion
of the effect of the
radiative corrections on the cosmological constant, see
e.g. ref.\cite{choi}.}.

The previous assumptions lead to some 
interesting relationships among the different soft terms,
with an automatical implementation of universality, something
which is phenomenologically welcome for FCNC reasons \cite{Dim81}.

The interest of this scenario raises two questions: 
\begin{description}

\item[\hspace{0.4cm}{\em i)}] Is there any
form of the superpotential $W(S)$ (preferably with some theoretical
justification) able to fulfill the previous
requirements (i.e. a minimum of the scalar potential at 
$\langle {\rm Re}S\rangle\simeq 2$ and $V=0$)? 
\item[\hspace{0.4cm}{\em ii)}] How good is expected to
be the tree-level approximation for $K(S,\bar S)$?

\end{description}

Regarding the first question, ({\em i}), we will present in sect.2
a simple analytical argument that proves that
, even if the potential
has a minimum at $\langle {\rm Re}S\rangle\simeq 2$ and $V=0$, there
must exist an additional minimum (or unbounded
from below direction) in the perturbative region of $S$-values for which
$V<0$. This can be proven for an arbitrary form $W(S)$. 
Similarly, we will show that the very existence of such a (local)
minimum is forbidden unless a really huge conspiracy of different
contributions to $W(S)$ takes place.

Concerning the second question, ({\em ii}), there are, unfortunately,
indications that the stringy non-perturbative corrections to the K\"ahler
potential may be sizeable \cite{Banksdine, Shenker}. In principle, 
this may be good news since 
it could help to avoid the above-mentioned problems. The trouble is that
very little is known about the form of these non-perturbative corrections.
On the other hand,
there are string arguments \cite{Banksdine} indicating that the stringy 
perturbative
and non-perturbative corrections to $W$ (and to the gauge kinetic function,
$f$, beyond one-loop) are negligible, so $W(S)$ must be dominated by
field-theory non-perturbative corrections (in particular gaugino condensation),
whose general form is known. In any case, symmetries plus analiticity
arguments strongly constrain \cite{Banksdine} the functional 
dependence of $W(S)$, which
must be $\sim\sum_i d_i e^{-a_iS}$.

{}From the previous reasons, it is suggestive to explore a generalized
 dilaton-dominated scenario in which $K(S,\bar S)$ is left as an {\em 
arbitrary} function, while $W(S)$ is assumed to be as above. 
In some sense, this precisely 
represents a philosophy opposite to the ``standard'' (tree-level)
dilaton-dominated scenario,
but in our opinion much more likely to be close to the actual facts. In
particular, it is interesting to analyze whether it is possible in this 
framework to avoid the shortcomings which are present in the tree-level
one. This scenario is investigated in sect.3, where generalized expressions
for the soft breaking terms are obtained.

Finally, it is tempting to examine the phenomenological
capabilities
of specific forms for the stringy non-perturbative contributions to $K$
suggested by theoretical work in the subject \cite{Shenker}. 
This will be done in sect.4

\section{Problems for the dilaton-dominated scenario}

Taking the tree-level expression for the K\"ahler potential
\be
\label{K}
K=-\log(S+\bar S)
+\hat K(T,\bar T,\phi_I,\bar \phi_I)\;\;,
\ee
where $T,\phi_I$ denote generically all the moduli and matter fields
respectively, the scalar potential in the dilaton-dominance assumption
reads
\be
\label{V}
V=\frac{1}{2\ {\rm Re}S}\left\{\left|(2\ {\rm Re}S)W_S-W\right|^2-3\left|
W\right|^2\right\}
\ee
with $W_S\equiv \partial W/\partial S$. If the scenario is realistic, 
the previous potential should  have a minimum at a realistic value
of $S$, say $S_0$, with
\be
\label{S2}
{\rm Re}S_0=\frac{1}{g^2}\simeq 2\;\;,
\ee
where, for simplicity of notation, $g$ denotes the gauge coupling constant
at the string scale ($\sim 10^{17}$ GeV). 
In addition, the cosmological constant should be vanishing, i.e.
$V(S_0)=0$. This implies
\be
\label{Lambda0}
\left|(2\ {\rm Re}S_0)W_S(S_0)-W(S_0)\right|=\sqrt{3}\left|
W(S_0)\right|
\ee
and, thus
\be
\label{ineq}
\frac{2\ {\rm Re}S_0}{\sqrt{3}+1}\leq 
\left|\frac{W}{W_S}\right|_{S=S_0}\leq
\frac{2\ {\rm Re}S_0}{\sqrt{3}-1}
\ee
Performing
the following change of variables 
\be
\label{change}
z=e^{-\beta S}\;\;\;(\beta\;{\rm arbitrary})
\ee
the physical region of $S$, i.e. ${\rm Re}S >0$, 
is mapped into the circle of radius 1 in the 
$z$-plane. The ``realistic minimum'' point, $z_0=e^{-\beta S_0}$, lies
somewhere inside the circle. In the new variable, the functions $W,W_S$
are written as
\bea
\label{Omegas}
W(S)&=&\Omega(z)
\nonumber\\
W_S(S)&=&\Omega_S(z)\equiv -\beta z \Omega '(z)
\eea
and condition (\ref{ineq}) becomes
\be
\label{ineq2}
\frac{2\ {\rm Re}S_0}{\sqrt{3}+1}\leq 
\left|\frac{\Omega(z)}{\Omega_S(z)}\right|_{z=z_0}\leq
\frac{2\ {\rm Re}S_0}{\sqrt{3}-1}
\ee
with ${\rm Re}S_0=-\log|z_0|/\beta$. Let us consider now the function
\be
\label{ro}
\rho(z)\equiv\frac{\Omega(z)}{\Omega_S(z)}\;.
\ee
Let us suppose for the moment that $\rho(z)$ is an analytical function 
with no poles inside the physical region $|z|<1$. Then, the maximum
of $|\rho(z)|$ in the region $|z|\leq|z_0|$ must necessarily occur (principle 
of maximum) at some point $z_M$ belonging to the boundary, namely 
the circle ${\cal C}\equiv\{|z|=|z_0|\}$.
If we consider the larger region
enclosed by the broader circle ${\cal C'}
\equiv\{|z|=|z_0'|\}$, with $|z_0'|>|z_0|$, the new maximum
of $|\rho(z)|$ must occur now at some point, say $z=z_1$, belonging 
to the boundary ${\cal C'}$. 
{}From (\ref{ineq2}) it is clear that at $z_1$
\be
\label{ineq3}
|\rho(z_1)|=\left|\frac{\Omega(z)}{\Omega_S(z)}\right|_{z=z_1}
> \frac{2\ {\rm Re}S_0}{\sqrt{3}+1}
\ee
Taking the radius of ${\cal C'}$ so that ${\rm Re}S_1\equiv
-\log|z_1|/\beta=\frac{\sqrt{3}-1}{\sqrt{3}+1}{\rm Re}S_0$,
we can write (\ref{ineq3}) as
\be
\label{ineq4}
\left|\frac{\Omega(z)}{\Omega_S(z)}\right|_{z=z_1}
=\left|\frac{W}{W_S}\right|_{S=S_1}
> \frac{2\ {\rm Re}S_1}{\sqrt{3}-1}
\ee
Therefore, at $S=S_1$ the potential (\ref{V}) has a negative value. On the other
hand $S_1$ still belongs to the perturbative region
\be
\label{S1}
{\rm Re}S_1\simeq 0.27 {\rm Re}S_0
\ee
which means $\alpha\simeq 0.14$. The only way-out to the previous argument
is to allow the function $\rho(z)\equiv\frac{\Omega(z)}{\Omega_S(z)}$ to
have some pole in the region enclosed by ${\cal C'}$. But then
$\left|\frac{\Omega}{\Omega_S}\right|\rightarrow\infty$ near the pole and
necessarily $\left|\frac{\Omega}{\Omega_S}\right|>
\frac{2\ {\rm Re}S}{\sqrt{3}-1}$ at some point with non-zero $\Omega$.
Hence we arrive to the same conclusion. 

\vspace{0.3cm}
\noindent
The previous argument shows that the realistic minimum assumed to take place
in the usual dilaton-dominated scenario can {\em never} correspond to a
global minimum. This is not certainly the most desirable situation.

\vspace{0.3cm}
\noindent
We can go a bit further and show that under very general assumptions,
the realistic point $S=S_0$ cannot correspond to $V(S_0)=0$.

{}From symmetry and analyticity arguments  we know \cite{Banksdine}
that the non-perturbative superpotential must take the form
$W=\sum_i d_i e^{-a_i S}$, so it is reasonable to assume that at the
realistic point ($S_0\sim 2$) $W$ is dominated by one of the terms, say
$W\sim e^{-a S}$ (as it happens for instance in usual gaugino 
condensation). Then, the vanishing of $\Lambda_{cos}$ at $S=S_0$,
eq.(\ref{Lambda0}), implies
\be
\label{conditiontree1}
\left(-a-\frac{1}{2{\rm Re}S_0}\right)^2
-\frac{3}{4({\rm Re}S_0)^2}=0\;.
\ee
{}From (\ref{conditiontree1}) we obtain $a\simeq (\sqrt{3}-1)/4$, absolutely
incompatible with the requirement of a hierarchically small SUSY breaking
(note that $\langle W\rangle\sim 1$ TeV requires $a\simeq 18$).

It could happen, however, that two o more terms of the form 
$W=\sum d_i e^{-a_i S}$
cooperate at the particular region $S\simeq 2$ to produce a more realistic
scenario\footnote{This is the mechanism of the so-called racetrack models to
generate SUSY breaking (see e.g. \cite{racetrack} and references therein).
These models, however, lead naturally to moduli dominance SUSY breaking rather
than dilaton dominance. For an attempt to generate dilaton dominance in
this way see ref.\cite{Halyo}}. 
It is interesting to show that for this
to happen in a dilaton dominated scenario 
a really huge conspiracy must take place. From eq.(\ref{ineq})
we see that the condition $V(S_0)=0$ implies
\be
\label{comp1}
|W_S| \sim |W|
\ee
Since $W_S=-\sum a_i d_i e^{-a_i S}$ and
the condition of hierarchically SUSY breaking requires 
$a_i\simgt O(10)$, it is clear that a cancellation between terms with
different exponents must occur inside $W_S$ for (\ref{comp1}) to be 
fulfilled. This can happen if the
$d_i$ coefficients are also (very) different, but is not enough: 
The extremalization condition $\partial V/\partial S=0$ applied to
the potential (\ref{V}) leads to two possible solutions,
\bea
\label{extr}
2\ {\rm Re}S\ W_S-W&=&0
\nonumber \\
(2\ {\rm Re}S)^2 W_{SS}&=&2W^*\frac{2\ {\rm Re}S W_S-W}{\left(
2\ {\rm Re}S W_S-W\right)^*}\;\;.
\eea
Since $(2\ {\rm Re}S\ W_S-W)$ is proportional to $F_S$, only the second
solution in (\ref{extr}) is compatible with a dilaton dominance 
scenario\footnote{The first solution of eq.(\ref{extr}) was shown in
ref.\cite{bru} to be incompatible with a global minimum at $V=0$, unless
a certain number of terms in the expansion $W=\sum d_i e^{-a_i S}$ were
relevant, in concordance with the general result obtained between
eqs.(\ref{Lambda0})--(\ref{S1}).}.
This requires
\be
\label{comp2}
(2\ {\rm Re}S)^2|W_{SS}| = 2|W|
\ee
This is an unnatural requirement since the typical size of $W_{SS}$
is $W_{SS}\sim a_i^2 W\gg W$. So, a second unpleaseant cancellation must ocurr
at $S=S_0$. Finally, one must demand that the extremum corresponds to a
minimum. The determinant of the Hessian matrix (evaluated in the variables
${\rm Re}S$, ${\rm Im}S$) reads
\be
\label{hess}
{\cal H}=-4\left\{(2\ {\rm Re}S)^2\left|
3W_{SS}W_S+W_{SSS}\left(2\ {\rm Re}S\ W_S-W\right)^*
\right|^2
-\frac{16}{(2\ {\rm Re}S)^4}|W|^4
\right\}
\ee
Using eqs.(\ref{Lambda0}, \ref{comp1}, \ref{comp2}), 
the condition for a minimum, ${\cal H}>0$, requires 
\be
\label{comp3}
\frac{2}{\sqrt{3}} ({\rm Re}S)^2|W_{SSS}|\sim |W_S|\sim |W|\;.
\ee
Again, this is an (even more) unnatural requirement, since typically
$|W_{SSS}|\sim a_i^3 |W|$, so a third fine cancellation must take place.

\vspace{0.3cm}
\noindent
To summarize the results of this section, the standard (tree-level) 
dilaton-dominated scenario
can {\em never} correspond to a global minimum of the potential at 
$V=0$. Similarly, under very general assumptions it cannot correspond
to a local minimum either, unless a really big conspiracy of different
contributions to $W(S)$ takes place. In addition, let us mention that
it has recently been
shown \cite{DL} that the effective low-energy scenario to which 
it gives place necessarily contains dangerous charge and color breaking 
minima.

\section{The generalized dilaton-dominated scenario}

The previous results, plus the fact that the K\"ahler potential is
likely to receive sizeable string non-perturbative corrections,
strongly suggest to consider a more general scenario, as
commented in the introduction. Thus, in this section we will study
how far we can go with the usual assumption
of ``dilaton--dominance'' (i.e. only $|F_S|\neq 0$), but leaving
the K\"ahler potential arbitrary. The potential reads
\be
\label{V2}
V=K_{S\bar S}\left| F_S\right|^2
-3e^K\left|W\right|^2\;,
\ee
where
\be
\label{Fsgen}
F_S=e^{K/2}\left\{
\left(K^{-1}\right)^{\bar \phi S}\left(
\partial_\phi W + W K_\phi\right)^*
\right\}
\ee
with $\phi$ running over all the chiral fields.
If we also assume vanishing cosmological constant, then
\be
\label{Fs}
\left| F_S\right|(K_{S\bar S})^{1/2}=\sqrt{3}e^{K/2}\left|W\right|=
\sqrt{3}m_{3/2}\;.
\ee
The passage to the effective low-energy theory involves a number of 
rescalings. In particular, the canonically normalized scalar and gaugino
fields are given by $\hat \phi_I=K_{I\bar I}\phi_I$, 
$\hat \lambda_a=({\rm Re}f_a)^{1/2}\lambda_a$ respectively, where
$K_{I\bar I }=\partial_{\phi_I}\partial_{\bar \phi_I}K$ and $f_a$ is
the gauge kinetic function. Likewise, the Yukawa couplings of the
effective superpotential are given by $\hat Y_{IJL}=
e^{K/2}(W^*/|W|)(K_{I\bar I}K_{J\bar J}K_{L\bar L})^{-1/2}Y_{IJL}$.
With these redefinitions the supersymmetric part of the Lagrangian is
obtained by applying the usual global-SUSY rules, while
the soft part of the Lagrangian is given by
\bea
\label{Lsoft}
-{\cal L}_{\rm soft}&=&
\frac{1}{2}M^a_{1/2}\bar {\hat \lambda_a} \hat \lambda_a +
\sum_I m_I^2|\hat \phi_I|^2  + 
\left( A_{IJL}Y_{IJL}\hat \phi_I\hat \phi_J\hat \phi_L
+ {\rm h.c.} \right)\; +\ \cdots \; .
\eea
The values of the gaugino masses, $M^a_{1/2}$, scalar masses,
$m_I^2$, and  coefficients of the trilinear scalar terms, 
$A_{IJL}$, can be computed using general formulae 
\cite{Kaplu93, Hall} and eqs.(\ref{V2}--\ref{Fs})
\be
\label{soft1}
\left|M_{1/2}\right|=\frac{1}{2}\sqrt{3}g^2 m_{3/2}(K_{S\bar S})^{-1/2}
\ee
\be
\label{soft2}
m_I^2=m_{3/2}^2\left[1-3K_{S\bar S}
\left(\frac{\partial_S\partial_{\bar S}K_{I\bar I}}{K_{I\bar I}}
-\frac{\left|\partial_S K_{I\bar I}\right|^2}{K_{I\bar I}^2}
\right) \right]
\ee
\be
\label{soft3}
\left|A_{IJL}\right|=\sqrt{3}m_{3/2}
\left(K_{S\bar S}\right)^{-1/2}\left|K_S+
\sum_{p=I,J,L}
\left(\frac{\partial_SK_{p\bar p}}{K_{p\bar p}}
\right) \right|\;.
\ee
We have assumed here that the Yukawa couplings appearing in the 
original superpotential, $W$, do not depend on the dilaton $S$. This is true
at tree-level (and thus at the perturbative level) and, since they are
parameters of the superpotential, they are not likely to be appreciably 
changed at the non-perturbative level \cite{Banksdine}.
The expression for the coefficient of the bilinear term, $B$, 
depends on the mechanism of generation
of the $\mu$ term, so we prefer to leave it as an independent parameter.
In the previous equations, besides the value of $m_{3/2}$, 
there are four unknowns: $K_S,K_{S\bar S},
\frac{\partial_SK_{I\bar I}}{K_{I\bar I}},
\frac{\partial_S\partial_{\bar S}K_{I\bar I}}{K_{I\bar I}}$. From the
phenomenological requirement of (approximate) universality \cite{Dim81} the
quantities $\frac{\partial_SK_{I\bar I}}{K_{I\bar I}},
\frac{\partial_S\partial_{\bar S}K_{I\bar I}}{K_{I\bar I}}$ should be
universal for all the $\phi_I$ (matter) fields. If we further assume
that $K_{I\bar I}$ does not get $S$-dependent contributions\footnote{
This, of course, may not occur. However, it is a common assumption
of all existing string-based models (including the usual dilaton-dominance
model). There are some examples (mainly orbifold compactifications) 
where the one-loop corrections to $K_{I\bar I}$ are known \cite{Kaplder}
and have been
incorporated in the analysis \cite{soft, Brignole}, but they are very small.} 
then there
are just two unknowns: $K_S,K_{S\bar S}$ (besides the value of $m_{3/2}$).
We obtain the following relations:
\bea
\label{relations}
m_I^2&=&m_{3/2}^2
\nonumber\\
\left|\frac{M_{1/2}}{m_{3/2}}\right|&=&\left[\frac{3g^4}
{4K_{S\bar S}}\right]^{1/2}
\nonumber\\
\left|\frac{M_{1/2}}{A}\right|&=&\left|\frac{g^2}{2K_S}\right|\;.
\eea
Notice that in the tree level limit $K_S=\frac{1}{2{\rm Re}S},\ K_{S\bar S}=
\frac{1}{4({\rm Re}S)^2}$, and we recover from eqs.(\ref{relations}) 
the usual tree level relations \cite{Kaplu93, Brignole}

Now, the non-perturbative superpotential must take the form
$W=\sum_i d_i e^{-a_i S}$, so, as explained above, 
it is reasonable to assume that at the
realistic point, $S\sim 2$, $W$ is dominated by one of the terms, say
$W\sim e^{-a S} $, as it happens for instance in usual gaugino 
condensation. Thus, the condition $\Lambda_{cos}=0$, i.e. eq.(\ref{Fs}), 
gives us a further constraint, namely
\be
\label{condition}
\left|-a+K_S\right|^2-3K_{S\bar S}= 0
\ee
which relates the values of  $K_S,K_{S\bar S}$ and $a$. Notice that 
the latter is essentially fixed by the condition of a hierarchical SUSY 
breaking ($a\simeq 18$), thus eqs.(\ref{relations}) and 
eq.(\ref{condition}) give a non-trivial scenario whose phenomenology
could be investigated.

\vspace{0.3cm}
\noindent
In obtaining eq.(\ref{condition}) from (\ref{Fs}) and (\ref{Fsgen}) 
we have ignored the possible dependence
of $W$ and $K$ on the $T$-field (or fields) as well as the mixing between
$S$ and $T$. This is justified by the following.
Working, for the sake of simplicity, with the dilaton $S$ and an overall
modulus $T$ the explicit form of $F_S$ is given by
\be
\label{Fs2}
F_S=e^{K/2}\left\{
\left(K^{-1}\right)^{\bar S S}\left(
\partial_S W + W K_S\right)^*
\ +\
\left(K^{-1}\right)^{\bar T S}\left(
\partial_T W + W K_T\right)^*
\right\}
\ee
The fact that, by assumption, $F_T=0$ at the minimum of the potential, implies
$\left(\partial_T W + W K_T\right)^*=\frac{\left(K^{-1}\right)^{\bar S T}}
{\left(K^{-1}\right)^{\bar T T}}
\left(\partial_S W + W K_S\right)^*$. Thus\\ 
\noindent
$F_S=
e^{K/2}\left[
\left(K^{-1}\right)^{\bar T T}\right]^{-1}\left|K^{-1}\right|
\left(\partial_S W + W K_S\right)^*=
e^{K/2}\left(K_{\bar S S}\right)^{-1}\left(\partial_S W + W K_S\right)^*$,
and the scalar potential (\ref{V2}) reads
\be
\label{V3}
V=e^K\left\{\left(K_{\bar S S}\right)^{-1}\left|\partial_S W + W K_S\right|^2
-3\left|W\right|^2
\right\}
\ee
where the explicit dependence on $T$ is irrelevant. Then, 
using $W\simeq e^{-a S}$
the condition $V=0$ translates into eq.(\ref{condition}) above.

\section{Ansatzs for non--perturbative K\"ahler potentials}

It is tempting to go a bit further in our analysis and explore the
phenomenological 
capabilities of explicit (stringy) non-perturbative effects on $K$
which have been suggested in the literature\footnote{For other attempts
in this sense, see ref.\cite{nilles}.}.

{}From the arguments explained in ref.\cite{Banksdine}, it is enough to focuss 
our attention on possible
forms of $K(S+\bar S)$, exploring the chances of getting a 
global minimum at $V=0$. From eq.(\ref{condition}) we can check that 
a successful generalized dilaton dominated scenario (i.e. with $V=0$ at
${\rm Re}S\simeq 2$) requires very sizeable 
non-perturbative corrections to $K$. This can be seen by considering the 
form of eq.(\ref{condition}) when $K$ is substituted by the tree level 
expression $K=-\log(S+ \bar S)$. Then, the left hand side of 
(\ref{condition}) reads
\be
\label{conditiontree}
\left[\left(-a-\frac{1}{2{\rm Re}S}\right)^2
-\frac{3}{4({\rm Re}S)^2}\right]_{{\rm Re}S\simeq 2},
\ee
which, as mentioned in sect.2, cannot be cancelled for any reasonable 
value of $a$.
The fact that $a\simeq 18$ implies that the non-perturbative ``corrections''
to $K_{S},K_{\bar S S}$ must indeed be bigger than the tree level values.
If we further demand that the potential has a minimum at $V=0$ this implies
\bea
\label{Conds}
\left(-a+\frac{1}{2}K'\right)K''-\frac{3}{4}K'''&=& 0
\nonumber\\
\left(-a+\frac{1}{2}K'\right)^2-\frac{3}{4}K''&=& 0
\eea
(at the realistic point ${\rm Re}S=2$), where the primes denote derivatives
with respect to ${\rm Re}S$ (the second equation is simply 
eq.(\ref{condition})). It would be nice if the previous conditions
could be fulfilled  by some simple form of $K$.

\vspace{0.3cm}
\noindent
Since in the known examples (mainly orbifold compactifications) the 
perturbative corrections to $K$ are small, we can temptatively write
$K({\rm Re}S)$ as
\be
\label{K2}
K({\rm Re}S)=-\log(2\ {\rm Re}S) + K_{np}({\rm Re}S)
\ee
where the first term corresponds to the tree level expression and $K_{np}$ 
denotes the non-perturbative contributions.

The next step is to choose some plausible form for $K_{np}({\rm Re}S)$ and to
study the form of $V$, see eq.(\ref{V3}). 
A reasonable assumption could be to impose that
$K_{np}({\rm Re}S)\rightarrow 0$ for 
${\rm Re}S\rightarrow\infty$ (i.e. the limit of 
vanishing gauge coupling) and that $K_{np}({\rm Re}S)$ 
is zero at all order in the
perturbative expansion. The field theory contributions to 
$K_{np}$ have been evaluated in ref.\cite{mariano}, but for a realistic
case they turn out to be too tiny to appreciably modify the
tree-level results\footnote{ 
More precisely, the authors of ref.\cite{mariano} obtain
$K=-3\log\left[(2\ {\rm Re}S)^{1/3}+e^{-K_p/3}+
de^{-(2\ {\rm Re}S)/2c}\right]$, where $K_p$ denotes the perturbative
corrections to the tree-level expression, $d$ is an unknown constant
and $c$ is essentially the coefficient of the one-loop beta function
(the non-perturbative superpotential goes as $e^{-3S/2c}$).
}. On the other hand, 
according to the work of ref.\cite{Shenker}, stringy
non-perturbative effects may be sizeable (even for weak four-dimensional
gauge coupling) and 
plausibly go as $g^{-p}e^{-b/g}$ with
$p,b\sim O(1)$. Then a simple possibility is to take
\be
\label{K3}
K_{np}=d\ g^{-p}e^{-b/g}
\ee
where $d,p,b$ are constants with $p,b>0$ and $g^{-2}={\rm Re}S$. 
Then, it can be explicitly shown 
that for $p=1$ the conditions (\ref{Conds}) can be satisfied for 
${\rm Re}S=2$ and $b\simeq 1$, leading indeed to the appearance of a minimum
with $V=0$. Unfortunately, the required value of $d$ is very large 
(and negative), which means that the model is not realistic, since 
$m_{3/2}^2=e^K|W|^2$ would be extremely small.

\vspace{0.3cm}
\noindent
Perhaps a more sensible approach is to write the decomposition (\ref{K2})
for the exponential of the K\"ahler potential, i.e.
\be
\label{K4}
e^{K}=e^{K_{tree}}+e^{K_{np}},
\ee
as is suggested by the fact that the SUGRA lagrangian has an exponential
dependence on $K$ (e.g. it can be written as 
${\cal L}=\left[e^{-K/3}\right]_D$), and then take
$e^{K_{np}}\sim d\ g^{-p}e^{-b/g}$.
(For small $K_{np}$ eqs. (\ref{K2}) and (\ref{K4}) would be
essentially equivalent.)

The numerical results indicate that, plugging this ansatz,
it is not difficult to get a minimum of the potential
using just one condensate with $a\simeq 18$ (thus guaranteeing the
correct size of SUSY breaking) and sensible values for the $d,p,b$ constants.
More precisely, for $d=-3$, $p=0$, $b=1$, there is a minimum of the potential
at ${\rm Re}S=2.1$. Unfortunately, the negative value of $d$ makes
this example unacceptable. Another example with positive $d$ is
$d=7.8$, $p=1$, $b=1$, which has a minimum at ${\rm Re}S=1.8$. However,
as a general result, playing just with these simple forms for $K$,
it seems impossible to get the minimum at $V=0$. Anyway, it is
impressive that just with one condensate and choosing very reasonable
values for $d$, $p$ and $b$ (note also that there is no fine-tuning in 
the previous choices), a minimum appears at the right value of the 
dilaton. Another typical characteristic of these examples is the appearance
of singularities in the potential caused by zeroes of the second derivative
of the K\"ahler potential, $K''$, at some particular values of $S$. 
Of course, this can be cured by additional terms in
$K$.

Finally, it would be worth to analyze the possible implications of 
this kind of non-perturbative K\"ahler potentials for other different matters,
such as the cosmological moduli problem \cite{moduli1, moduli2}

\section{Conclusions}

We have shown that the usual dilaton dominance scenario, derived
from the tree level K\"ahler potential, can {\em never} correspond to 
a global minimum of the potential at 
$V=0$. Similarly, it cannot correspond
to a local minimum either, unless a really big conspiracy of different
contributions to the superpotential $W(S)$ takes place.

These results, plus the fact that the K\"ahler potential is
likely to receive sizeable string non-perturbative corrections,
strongly suggest to consider a more generalized scenario, leaving the 
K\"ahler potential arbitrary. In this way we obtain generalized
expressions for the soft breaking terms. A predictive scenario arises
if one further assumes, as it is usually done, that the matter fields
kinetic terms, $K_{I \bar I}$, do not get sizeable $S$-dependent
contributions and that the non-perturbative superpotential
$W=\sum_i d_i e^{-a_i S}$ is dominated at the
realistic point by one of the terms, say
$W\simeq e^{-a S}$ (as it happens for instance in usual gaugino 
condensation).
Then, the relevant formulae for the scalar and gaugino masses,
$m_I^2, M_{1/2}$, and for the coefficient of the trilinear scalar terms,
$A$, are given (see eqs.(\ref{relations}, \ref{condition})) by
\bea
\label{relations2}
m_I^2&=&m_{3/2}^2
\nonumber\\
\left|M_{1/2}\right|&=& m_{3/2}\frac{3 g^2}
{|K'-2a|}
\nonumber\\
\left|A\right|&=&\left|M_{1/2}\right|
\left|\frac{g^2}{K'}\right|\;,
\eea
where $K'\equiv\partial K/\partial ({\rm Re}S)$ and $a\simeq 18$ 
from the condition of hierarchical SUSY breaking.

We have explored the phenomenological capability of some
theoretically motivated forms of non-perturbative K\"ahler potentials, 
showing that for reasonable
choices it is easy to get a minimum of the potential at the 
realistic value $S\sim 2$ with just one condensate and with no fine-tuning
at all. The goal of getting a vanishing cosmological constant remains
however as a more challenging task.

Finally, it would be nice if the new string-duality techniques (see in
particular ref.\cite{w}) could provide more precise information about
the form of $K$, so its phenomenological viability could be tested in
the sense exposed in this paper.

\section*{Acknowledgements}
I thank M. Dine and M. Quir\'os for very useful discussions.



\def\MPL #1 #2 #3 {{\em Mod.~Phys.~Lett.}~{\bf#1}\ (#2) #3 }
\def\NPB #1 #2 #3 {{\em Nucl.~Phys.}~{\bf B#1}\ (#2) #3 }
\def\PLB #1 #2 #3 {{\em Phys.~Lett.}~{\bf B#1}\ (#2) #3 }
\def\PR  #1 #2 #3 {{\em Phys.~Rep.}~{\bf#1}\ (#2) #3 }
\def\PRD #1 #2 #3 {{\em Phys.~Rev.}~{\bf D#1}\ (#2) #3 }
\def\PRL #1 #2 #3 {{\em Phys.~Rev.~Lett.}~{\bf#1}\ (#2) #3 }
\def\PTP #1 #2 #3 {{\em Prog.~Theor.~Phys.}~{\bf#1}\ (#2) #3 }
\def\RMP #1 #2 #3 {{\em Rev.~Mod.~Phys.}~{\bf#1}\ (#2) #3 }
\def\ZPC #1 #2 #3 {{\em Z.~Phys.}~{\bf C#1}\ (#2) #3 }

\end{document}